# Comparison of gait phase detection using traditional machine learning and deep learning techniques

Farhad Nazari*, *Member*, IEEE, Navid Mohajer, Darius Nahavandi, *Member*, IEEE,
and Abbas Khosravi *Senior Member*, IEEE
*Institute for Intelligent System Research and Innovation (IISRI), Deakin University, Australia*

*Abstract*—Human walking is a complex activity with a high level of cooperation and interaction between different systems in the body. Accurate detection of the phases of the gait in real-time is crucial to control lower-limb assistive devices like exoskeletons and prostheses. There are several ways to detect the walking gait phase, ranging from cameras and depth sensors to the sensors attached to the device itself or the human body. Electromyography (EMG) is one of the input methods that has captured lots of attention due to its precision and time delay between neuromuscular activity and muscle movement. This study proposes a few Machine Learning (ML) based models on lower-limb EMG data for human walking. The proposed models are based on Gaussian Naive Bayes (NB), Decision Tree (DT), Random Forest (RF), Linear Discriminant Analysis (LDA) and Deep Convolutional Neural Networks (DCNN). The traditional ML models are trained on hand-crafted features or their reduced components using Principal Component Analysis (PCA). On the contrary, the DCNN model utilises convolutional layers to extract features from raw data. The results show up to 75% average accuracy for traditional ML models and 79% for Deep Learning (DL) model. The highest achieved accuracy in 50 trials of the training DL model is 89.5%.

## I. Introduction

Walking is a complex human activity that involves a high level of cooperation and interaction between the central nervous system, bones and muscles in the human lower limb, happening in a cycle known as the gait cycle [1]. Precise and real-time detection of the gait phase is crucial for accurately controlling a lower-limb assistive device, e.g. exoskeleton [2], prostheses or other wearable devices [3]. The Ankle Mimicking Prosthetic Foot 3 is an example of using gait phase detection to control the motor at the ankle [4]. A gait cycle starts with the heel strike of one leg (Initial Contact or IC) and finishes with the heel strike of the same leg [5]. Partitioning the gait cycle can be done in different ways. Perhaps the most popular approach in the literature is to divide it into stance and swing phases [6]. Some applications require dividing the stance phase into two mid-stances [7]. Some other studies reported six distinct phases for the gait cycle [8].

Various input methods can be used to determine the gait phase depending on the application. Some indoor applications use non-wearable sensors like cameras [9] or ultrasonic sensors [10]. However, most assistive devices use wearable sensors like accelerometers [11], inertial measurement units (IMUs) [12], gyroscopes [13], force myography [14], footswitches [15] joint angular [16] and surface electromyography (EMG) sensors [17]. Each sensor comes with some pros and cons. While EMG sensors benefit from the time delay between the neuromuscular signal and muscle movement, kinetic [18], [19] and kinematic [20] sensors can be embedded into the assistive device and do not need to be directly attached to the user's skin. The other benefit of EMG based methodologies is that they can be more practical for the weak, elderly and people with disability. As long as the patient can produce neuromuscular signals, they can be detected by EMG sensors [21], and there is no need for actual limb movement to enable assistive devices.

Different methodologies can be implemented to detect the gait phase from the input signals. Some of them work based on threshold selection like zero-crossing detection [22], and dynamic variations [23], gait symmetry [24], fast Fourier transform [25], wavelet transform [26], etc. Other approaches based on Machine Learning (ML) techniques and using manually extracted features are gaining popularity in recent years. Some of these methods are Linear Discriminant Analysis (LDA) [27], Support Vector Machines (SVM) [17], Gaussian Mixture Model (GMM) [28] and Decision Tree (DT) [29]. Principal Component Analysis (PCA) is another widely used technique to reduce the dimensionality of the input data [30], which is especially helpful in reducing the computation cost for real-time applications.

In recent years, with the increase in computation power and data availability, Artificial Neural Networks (ANN) like Shallow Learning [31] and Deep Learning (DL) [32] techniques have gained popularity. Multi-layer structures of DL networks help extract higher-level features from the input signal. Examples of these networks are Multi-Layer Perceptron (MLP) [33], Convolutional Neural Networks (CNN) [34], Long Short Term Memory (LSTM) [35] and combined ones like CNN-LSTM [36]. However, the number of trainable parameters in these models is huge and therefore requires lots of training data to train all these parameters and develop an accurate model accurately. Since these models are based on supervised learning methods, they require labelled data for training. This can limit the application as data must be collected in different controlled scenarios [34]–[36]. Alternatively, this study uses mathematical models to extract the desired labels from other relevant information existing in the data [37], allowing to utilise data that has been collected for other purposes. The broad adaptation of these methods can bypass the data availability problem and develop more sophisticated models to be implemented in human assistive devices.

The other limitation of these systems is their reliance on the quality of physiological signals like EMG. As stated earlier, these signals come with many advantages for assistive devices. However, the quality of their signals can be affected by so many internal and external factors. What is more, surface EMG sensors can become loose during the motion, reducing the signal to noise ratio to near zero. This study proposes some ML models utilising multi-channel signals. Lower-limb EMG data has been used to predict the gait phase. Channels with the corrupted recording (almost pure noise) have not been removed to represent the real-life possibility of signal corruption. In the next section, after discussing the dataset, the data processing and feature extraction will be illustrated. Then the proposed models and methodologies will be described in section III. Section IV discusses the results, followed by the last section, the conclusion.

* corresponding author:  f.nazari@deakin.edu.au

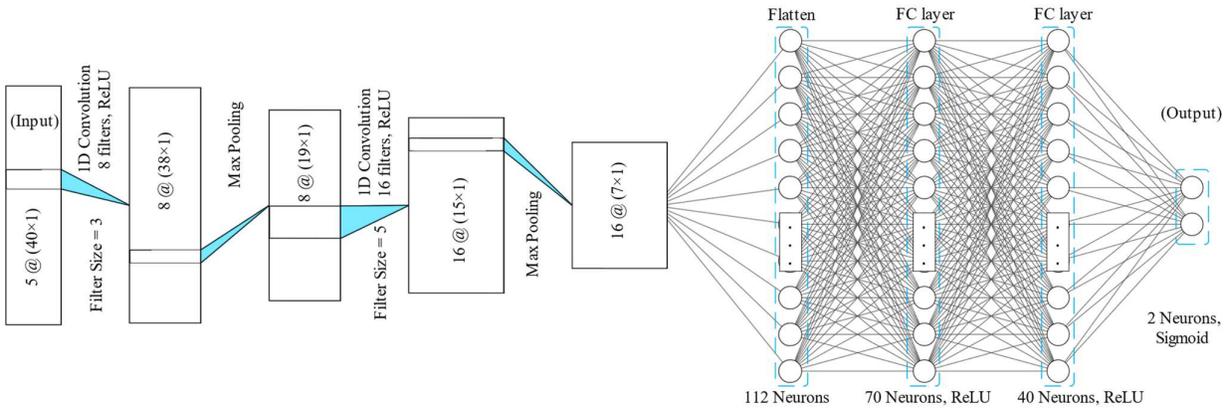

Fig. 1. The structure of the Deep learning model. The schematic has been created using alexlenail.me.

## II. Data Analysis

### A. Dataset

The lower limb muscle activation EMG dataset from Mohr et al., which is publicly available at the Mendeley Data repository, has been used in this study [38]. The dataset contains EMG signals from five muscles from each leg as well as heel strikes from 61 participants walking for 49 gait cycles. The signals have been recorded from Biceps Femoris (BF), Vastus Lateralis (VL), Medial Hamstrings (MH), Gastrocnemius Medialis (GM) at 1500Hz.

### B. Data Processing

First, heel strikes of both dominant and non-dominant legs have been studied to find any gait duration and pattern abnormalities. The first and last cycles of the recorded data for all subjects and the entire data of subjects 151 and 176 have been removed from the study. The reason for this was too much variation in the gait period or the difference between the first and second (standard deviation of more than 20% of the mean value). The gait phase has been calculated from heel strikes for each leg, based on [37]. These extracted labels have been added to the EMG signals of the corresponding leg.

EMG signals outside 20 and 300Hz have been filtered using a fourth-order two-sided Butterworth filter, then downsampled to 500Hz. Each channel has been mapped to a mean value of zero and a standard deviation of one to minimise the effect of signal quality in each sensor and person. The data has been transformed into 3D tensor for algorithms with 80ms window size and 32ms sliding between them. In another word, each window has 48ms overlap with the previous one. The outcome was a 443668×40×5 tensor. This tensor will be used to train and test the proposed models.

Tab 1. variance importance of the first five components

| Principal Component | PC1 | PC2 | PC3 | PC4 | PC5 |
|---|---|---|---|---|---|
| Variance Ratio | 0.35 | 0.21 | 0.11 | 0.07 | 0.06 |

### C. Hand-crafting features

Directly using large raw EMG data for classification usually decreases the model's efficiency [39]. Hand-crafting features from EMG signal, known as feature engineering, shrinks the size while maintaining important information, which improves the classifier's performance [36]. This is especially important for real-time prediction. Hand-crafting features can be done in time-, frequency-, or time-frequency domains [40]. Since time-domain features are computationally cheaper to calculate [41], they gained popularity in engineering and medical practices [40]. Variance (VAR), Mean Absolute Value (MAV), Root Mean Square (RMS) and Integrated EMG (IEMG) are some of the most popular feature extraction methods. Other feature engineering methods are Average Amplitude Change (AAC), Waveform Length (WL) and Difference Absolute Standard Deviation Value (DASDV). Frequency-based features are Sign Slope Change (SSC), MYOPulse Percentage (MYOP), Zero Crossing (ZC) and Wilson Amplitude (WAMP). Fast Fourier Transform is one of the most popular methods to transform the signal from the time domain spectrum into the frequency domain [42]. Mean power frequency (MPF) mean frequency (MNF), and median frequency (MDF) are other popular frequency-domain features.

In this study, we extracted four features from each channel time-series data to use them as the input to the algorithms with, resulting in an input size of 20 [43]. This can also minimise the stochastic effect of raw EMG data on our models, as traditional ML algorithms may suffer picking patterns from volatile and noisy data.

The four extracted features for this study are: standard deviation (σ), mean absolute value (MAV), zero crossing (ZC), and mean absolute deviation (MAD). These features are among the most widely used ones in the literature while they do not directly correlate with each other (like variance and standard deviation).

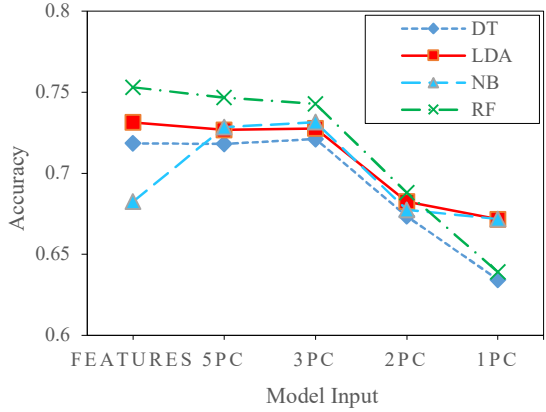

Fig. 2. The average accuracy of ML algorithms over 50 trials using different inputs.

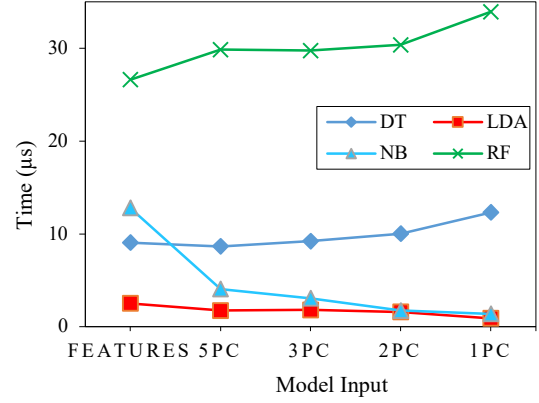

Fig. 3. The average computation cost of each ML model on different input types.

Equations one to four show the calculation of features. After extracting these features from each window section, we passed the results through a standard scalar which maps the data to zero mean and unit variance.

$$ZC = \frac{1}{2}\sum_{i=1}^{n-1} sgn(x_i \times x_{i+1}) \times u(|x_i - x_{i+1}| - \theta) \quad (1)$$

$$MAV = \frac{1}{n}\sum_{i=1}^{n}|x_i| \quad (2)$$

$$\sigma = \sqrt{\frac{\sum_{i=1}^{n}(x_i - \bar{x})^2}{n}} \quad (3)$$

$$MAD = \frac{1}{n}\sum_{i=1}^{n}|x_i - \bar{x}| \quad (4)$$

where:

n is the number of values in each window,
sgn is the sign function,
u is the unit step function,
$x_i$ is the dataset value,
$\bar{x}$ is the population mean,
θ is the threshold value; 3% MAV of the EMG channel.

### D. PCA

The features extracted in the previous section were reduced to its main components using PCA. Tab 1 shows the variance importance of the first five principal components.

### III. METHOD

We trained different ML models on random portions of EMG signals to evaluate the possibility of gait phase detection from lower-limb muscles (VL, BF, MH, GL, GM). Because of the 20-80ms time delay between neuromuscular activity and limb motion, a window size in this range can potentially help with real-time gait detection. In this research, 80ms window size has been chosen to be short enough for real-time computation and long enough to contain meaningful information.

Each model has been trained and tested 50 times using a random 90% of the data for training and 10% for testing. The train and test dataset has been separated based on subjects to prevent data leakage and eliminate the effect of learned patterns related to individual gait patterns. This also ensures a balanced left and right leg data split into the test and training sets. The results have been recorded and averaged out after 50 trials.

### A. Traditional ML Algorithms

The hand-crafted features in the previous section have been used to train Gaussian Naive Bayes (NB), Decision Tree (DT), Random Forest (RF) and Linear Discriminant Analysis (LDA) models. The hyperparameters of models have been optimised in each trial based on the training dataset and random search method. After training the optimal model, it was evaluated on the test dataset while the performance time was recorded. The same methodology has been used to develop models based on the same ML algorithms, with the input being reduced data derived from PCA. For this purpose, models have been trained on one, two, three and five principal components. Training and testing of the mentioned models have been completed on an Intel Core i9-11900F CPU. The CPU at no point was throttling.

### B. Deep Convolutional Neural Networks

A new deep learning model has been developed to improve performance, get closer to the limits of the proposed method, and use it to compare and evaluate traditional ML models. For this purpose, we let the DL model extract the useful features from the filtered and downsampled data rather than manually extracting features. Two layers of the 1D convolution layer were responsible for extracting features from 80ms portions of each EMG channel. A MaxPooling layer followed each layer to pick the most important features. The results were fed into two fully connected layers to find the deeper patterns in all five channels of the signals. After each layer, a heavy dropout of 30% was applied to prevent overfitting. The output layer is also a dense layer with two neurons to predict the gait phase. Fig 1 shows the structure of the deep learning model.

### IV. RESULTS AND DISCUSSION

In each training round, all models have been tuned and trained on the same set of training and test subjects to make the comparison between models easier. For the same reason, the initial random weights of the DL model have been saved and reused in each trial.

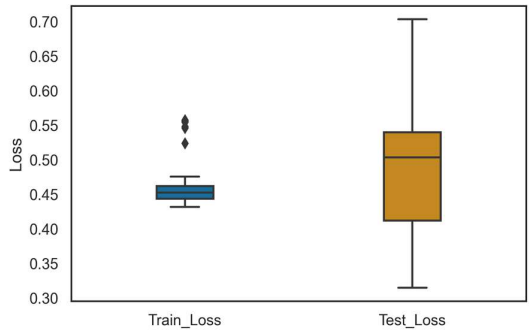

Fig. 4. The spread of train and test loss of the proposed DL model over 50 trials.

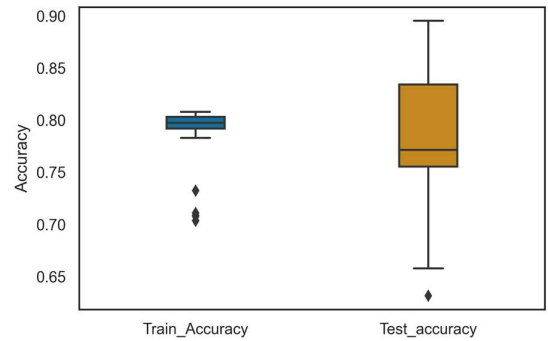

Fig. 5. The spread of train and test accuracy of the proposed DL model over 50 trials.

### A. Traditional ML models

Fig. 2 shows all models' average accuracy, with the inputs being the extracted features or their principal components. Random Forrest showed the highest accuracy, with its input being the hand-crafted features or its five or three Principal Components (PC). The next good performing model was the Gaussian Naive Bayes model with three PC inputs followed by LDA with its inputs being similar to the RF. Finally, the weakest performance with reduced data was the Decision Tree.

Regarding the effect of input type on performance, for RF and LDA, reducing the dimensionality of the data up to 3PC reduces the accuracy insignificantly but drops sharply afterwards. The DT model shows similar behaviour except for delivering the highest accuracy at 3PC. Reducing the dimensionality increases the accuracy for the NB model until 3PC and drops afterwards. Overall, reducing the data size to only three of its main components (67% of variance) seems to maintain a lot of information as it does not harm the performance in a meaningful way.

Regarding performance time, LDA was the fastest model, followed by NB. RF was the most computationally expensive model with an average of 3ms. DT, on average, took around 1ms to make its predictions. In LDA and NB models, reducing dimensionality reduces the computation time, which we were expecting. However, DT and RF show opposite behaviour. This can result from the change in the measure function or an increase in sample split or depth of the models. In other words, more sophisticated models are required to predict the gait phase from inputs with reduced dimensionality.

### B. DL model

In each trial of training and testing the proposed deep convolutional neural networks model, the training has been continued till there was no improvement in the validation loss for 100 epochs. However, the best performing model has been called back based on test accuracy rather than loss-value. Figs 4 shows the sparse categorical cross-entropy loss value spread for the DL model over 50 trials. Similarly, fig 5 shows the test accuracy spread for training and testing. The DL model showed average accuracy of 0.79 beating the best performing RF by 4%, NB and LDA by 6% and DT by 7%.

Considering the computation cost of neural networks, only 4% lower accuracy of RF model on hand-crafted features or 5% on three principal components seems good, and shows the acceptable performance of traditional ML models. On the other hand, the significant variation in test accuracy in different DL trials proves the inherent noise or inconsistency in the data. The maximum achieved accuracy in all trials was 89.5%, with the minimum being 0.63. This means that the data corresponding to some subjects is noisier than the others. Depending on which ones become test subjects, the results will be different. Another witness to this hypothesis is the relation between train and test accuracy.

As shown in Fig 6, for trials that resulted in low test accuracy, the test to train acc ratio is well below one, and the model demonstrates much higher training accuracy compared to the testing performance. This is contrary to the high test accuracies where the ratio is well above one, showing much higher test performance compared to the training. Both observations demonstrate an imbalance noise in the dataset that is not equally distributed among all subjects. We hypothesise that due to the significant portion size of the training data in each trial (around 90%), it is highly likely that the trained models have experienced distorted data to some degree, harming the performance. In contrast, the test dataset covers only around 10% of the whole data and is more likely to be affected by distorted data with more variation, resulting in more variation in testing accuracy. This is also visible in train and test accuracy and loss in figs 4 and 5.

Our first assumption for this distortion in data was the data corresponding to the participants with knee injury history, as this data was initially collected for pathological purposes. However, we did not witness any performance improvements after removing all subjects with a history of a knee injury. By identifying and removing distorted data, one can probably achieve the performance of higher the ones reported in this study. Because EMG signals are intuitive to understand, this is not easy to achieve using traditional preprocessing methods. One possible solution could be to train a NN model to help with distorted data identification.

### V. CONCLUSION

Gait phase detection is crucial for enabling the control system of human assistive devices, e.g. prostheses and exoskeletons. Among all methodologies to detect the gait phase, using Electromyography (EMG) has attracted so much attention due to its precision, time delay between the neuromuscular activity and muscle movement and ability to be implemented for weak, old and people with disability.

This study proposed and evaluated various models to interpret the EMG signals and detect the gait phase. We presented some traditional ML-based and Deep Convolutional Neural Networks (DCNN) models. Traditional models have been developed using hand-crafted features or their reduced principal components, whereas the DCNN model has been trained on filtered downsampled signals. Among proposed models, Random Forrest show satisfactory results and Gaussian Naive Bayes, Linear Discriminant Analysis and Decision Tree show acceptable performance. The average DCNN model accuracy was just 4% more than the best performing ML model. However, 26.5% variation in the accuracy results in the DCNN models show the potential of some distortion in the EMG data that was not easy to detect in the preprocessing stage. For the future work, we will consider complementary models to detect the distorted data and minimise their effect on our models.

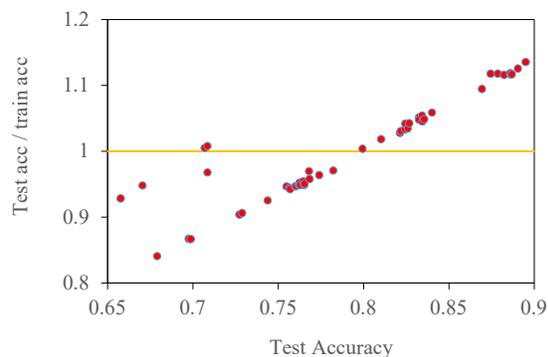

Fig. 6. The test accuracy to train accuracy ratio for different achieved test accuracies.

For future work and directions, we suggest to implementing and asses these methods on simalution platforms [44] and robotics [45], [46]. Furthermore, they can be paired with haptic feedback [47] and emotion recognition methodologies [48], body position [49], as well as autonomous control and learning systems [50]–[52] build a human-in-the-loop system.